\documentclass[twoside,twocolumn,english,aps,pre,showpacs]{revtex4}
\usepackage[T1]{fontenc}
\usepackage[latin1]{inputenc}
\usepackage{amsmath}
\usepackage{babel}
\usepackage{graphics}
\usepackage{amssymb}

\makeatletter

\providecommand{\LyX}{L\kern-.1667em\lower.25em\hbox{Y}\kern-.125emX\@}

\makeatother
\begin{document}
\pacs{61.43.Bn, 61.43.Fg, 02.30.Nw, 02.70.Hm}

\title{Instantaneous frequency and amplitude identification using wavelets:
Application to glass structure}

\author{J.D. Harrop, S.N. Taraskin, and S.R. Elliott}

\affiliation{\emph{Department of Chemistry, University of Cambridge, Cambridge
CB2 1EW, UK.}}

\date{15 February 2002}

\begin{abstract}
This paper describes a method for extracting rapidly varying, superimposed
amplitude- and frequency-modulated signal components. The method is
based upon the continuous wavelet transform (CWT) and uses a new wavelet
which is a modification to the well-known Morlet wavelet to allow
analysis at high resolution. In order to interpret the CWT of a signal
correctly, an approximate analytic expression for the CWT of an oscillatory
signal is examined via a stationary-phase approximation. This analysis
is specialized for the new wavelet and the results are used to construct
expressions for the amplitude and frequency modulations of the components
in a signal from the transform of the signal. The method is tested
on a representative, variable-frequency signal as an example before
being applied to a function of interest in our subject area - a structural
correlation function of a disordered material - which immediately
reveals previously undetected features.
\end{abstract}
\maketitle

\section{Introduction}

Although the first example of a wavelet basis dates back to 1910~\cite{Haar1910},
it was not until the early 1980's, with the work of Goupillaud, Morlet
and Grossmann~\cite{Goupillaud1984} in seismic geophysics, that
the wavelet transform (WT) became a popular tool for the analysis
of signals with non-periodic characteristics, termed \emph{non-stationary}
signals~\cite{vandenBerg1999}.

The WT allows a signal to be examined in both the time- and frequency-domains
simultaneously. The WT as a time-frequency method has replaced the
conventional Fourier transform (FT) in many practical applications.
The WT has been successfully applied in many areas of physics~\cite{vandenBerg1999}
including astrophysics, seismic geophysics, turbulence and quantum
mechanics, as well as many other fields including image processing,
biological signal analysis, genomic DNA analysis, speech recognition,
computer graphics and multifractal analysis.

The term WT is conventionally used to refer to a broad selection of
transformation methods and algorithms. In all cases, the essence of
a WT is to expand the input function in terms of oscillations which
are localized in both time and frequency.

Different applications of the WT have different requirements. Image
compression, for example, often uses the discrete WT (DWT) to transform
data to a new, orthogonal basis set where the data are hopefully presented
in a more redundant form~\cite{Prasad1997}. Other applications,
particularly signal analysis~\cite{Silverman2000}, use the CWT,
sacrificing orthogonality for extra precision in the identification
of features in a signal.

The principal aim of this paper is to present the WT in a form well
suited to the analysis of one-dimensional signals whose frequency
components have rapidly varying frequency and amplitude modulations.
In order to achieve this aim, we introduce a new `tunable', complex
wavelet. This wavelet is based upon the well-known Morlet wavelet~\cite{Goupillaud1984}
but is better suited to high-resolution analysis. The features of
the WT using the proposed wavelet are understood through an asymptotic
stationary-phase approximation to the integral expression of the WT
specialized to the new wavelet. We demonstrate the properties of the
WT using two example functions, a mathematical function and the other
a realistic physical model function.

We are interested in exploiting the complex WT in the analysis of
structural correlation functions which describe the atomic structure
of disordered materials. As these correlation functions have different
spatial regimes, they may be classed as non-stationary signals. Despite
the overwhelming success of the WT in other fields, we are aware of
only one other paper on this application of the WT. In that application,
Ding et al.~\cite{Ding1998} studied an experimentally observed correlation
function of vitreous silica using the Mexican Hat wavelet. We improve
upon this single, prior application in three significant ways, namely
the use of the complex WT, a `tunable' wavelet and the method of interpreting
the resulting transforms.

We then analyze the reduced radial distribution function (RRDF) of
a structural model of a one-component glass with pronounced icosahedral
local order~\cite{Dzugutov1993}. The resulting WT clearly shows
the existence of different frequency components in the RRDF and their
exponential decay. These features were not clearly detectable by earlier
methods (c.f.~Ref.~\onlinecite{Ding1998}).

In Sec.~\ref{Sec: Formulation} we review the mathematical framework
of the wavelet transform and discuss some mother wavelet functions
before modifying an existing wavelet for our purposes. In Sec.~\ref{Sec: Instantaneous Frequency and Amplitude}
we consider the WT of a general oscillatory signal using the new wavelet.
The results are then used to interpret the wavelet transforms of a
variable-frequency example function in Sec.~\ref{Sec: Example Function}
and of the RRDF in Sec.~\ref{Sec: Reduced Radial Distribution Function}.
Concluding remarks can be found in Sec.~\ref{Sec: Conclusions}.

\section{Formulation\label{Sec: Formulation}}

The underlying WT used in this paper can be completely described as
a one-dimensional complex, continuous WT using wavelets of constant
shape~\cite{Daubechies1992}. We begin by examining the formulation
of this WT in terms of an integral transform, before examining the
choice of mother wavelet function.

For simplicity we use time-frequency terminology, considering the
signal to be an input function of time, \( f(t) \). The CWT is an
integral transformation which expands an input function \( f(t) \)
in terms of a complete set of basis functions \( \xi (t;a,b) \).
These basis functions are all the same shape as they are defined in
terms of dilation by \( a \) and translation by \( b \) of a mother
wavelet function \( \psi (t) \):\begin{equation}
\label{Eqn: Dilation and translation of the mother wavelet}
\xi (t;a,b)=|a|^{-1}\psi \! \left( \frac{t-b}{a}\right) ,
\end{equation}
with \( a,b\in \mathbb {R} \) and \( a\neq 0 \).

The CWT \( F(a,b) \) is defined as the inner product:\begin{equation}
\label{Eqn: Integral expression for the wavelet transform}
F(a,b)=\left\langle \xi |f\right\rangle \equiv |a|^{-1}\int _{-\infty }^{\infty }\psi ^{*}\! \left( \frac{t-b}{a}\right) \, f(t)\, dt.
\end{equation}

The original formulation of the CWT~\cite{Goupillaud1984} used a
prefactor \( |a|^{-\frac{1}{2}} \) to give a normalization to unity,
\( \left\langle \xi |\xi \right\rangle =1 \)). We choose an alternative
prefactor \( |a|^{-1} \) (giving \( \left\langle \xi |\xi \right\rangle =|a|^{-1} \))
after Delprat et al.~\cite{Delprat1991}. As we shall see, this formulation
of the CWT allows for simple frequency identification by examining
the maxima in the modulus of the CWT with respect to the scale \( a \).

In order to understand the CWT, it is useful to relate it to the FT.
The FT has a non-localized, plane-wave basis set and, therefore, has
a single transform parameter - the frequency \( \omega  \). In contrast,
the basis set of the CWT contains localized oscillations characterised
by two transform parameters - the scale (or dilation) \( a \) and
the translation (or position) \( b \). It is this critical difference
which makes the CWT preferable for the analysis of non-stationary
signals.

We are free to choose a functional form for \( \psi (t) \), subject
to some constraints. Some of these constraints are forced upon us
whereas others arise from the practical usefulness of the resulting
transform.

In order to recover a function from its wavelet transform via the
\emph{resolution of the identity}~\cite{Daubechies1992}, \( \psi (t) \)
must satisfy an \emph{admissibility condition}. Although we do not
make direct use of the resolution of the identity in this paper, we
require that our choice of \( \psi (t) \) satisfies this condition
to ensure that all information about the signal is retained by the
transform. The admissibility condition is essentially that the FT,
\( \hat{\psi }(\omega )=\left\langle e^{i\omega t}|\psi \right\rangle  \),
satisfies the relation \( \hat{\psi }(0)=0 \), equivalent to requiring
that the mother wavelet and, hence, the basis wavelets, have a mean
of zero.

Beyond simply satisfying the admissibility condition, it is practically
useful to create mother wavelet functions which mimic features of
interest in the signal. In the case of time-frequency analysis, mother
wavelet functions are chosen which represent localized sinusoidal
oscillations. The resulting wavelet transforms can then be used to
extract instantaneous measures of frequency and amplitude. The uncertainty
principle dictates that the product \( \Delta t\Delta \omega  \)
of the time and frequency uncertainties of such wavelets has a lower
bound. It is no surprise, therefore, that this class of mother wavelet
functions are typically based upon Gaussians. However, it is still
possible to trade temporal precision for frequency precision by altering
the number of oscillations in the envelope of the mother wavelet.

The simplest such wavelet is the {}``Mexican Hat'' wavelet which
mimics a single oscillation and is commonly used in signal analysis.
The functional form of this wavelet is the second derivative of a
Gaussian. This wavelet offers good localisation in the time domain
whilst retaining admissibility. However, this wavelet has two major
drawbacks for general signal analysis: (i) useful information can
only be extracted from the WT at discrete intervals where the wavelets
are in phase with the signal, and (ii) the time-frequency resolution
is fixed.

The former drawback has been overcome by the invention of complex
wavelets which mimic localized plane waves. The WT can be computed
separately for the real and imaginary parts, yielding a complex scalar
field, \( F(a,b) \), where the modulus and argument of \( F \) represent
the amplitude and phase of the signal, respectively.

The latter drawback has been overcome by the invention of tunable
wavelets which include an additional parameter to the mother wavelet
function controlling the number of oscillations in the envelope.

Goupillaud, Morlet and Grossman overcame these problems simultaneously
with the invention of a modulated Gaussian wavelet, now known as the
{}``Morlet'' wavelet~\cite{Goupillaud1984}. This wavelet has a
parameter, \( \sigma  \), which controls the number of oscillations
in the envelope, allowing time and frequency uncertainties to be traded.
Thus the Morlet wavelet can be expressed as:\begin{eqnarray}
\psi _{\textrm{M}}(t;\sigma ) & = & \pi ^{-\frac{1}{4}}c_{\textrm{M}}(\sigma )\, e^{-\frac{1}{2}t^{2}}\left( e^{i\sigma t}-\kappa (\sigma )\right) ,\label{Eqn: Morlet wavelet} 
\end{eqnarray}
where \( c_{\textrm{M}}(\sigma )=(1-2e^{-\frac{1}{4}\sigma ^{2}}\kappa (\sigma )+\kappa ^{2}(\sigma ))^{-1/2} \)
and the parameter \( \kappa (\sigma ) \) allows the admissibility
condition to be satisfied.

The FT of this wavelet is:\begin{eqnarray}
\hat{\psi }_{\textrm{M}}(\omega ;\sigma ) & = & \pi ^{-\frac{1}{4}}c_{\textrm{M}}(\sigma )\left( e^{-\frac{1}{2}\left( \omega -\sigma \right) ^{2}}-\kappa (\sigma )\, e^{-\frac{1}{2}\omega ^{2}}\right) .\label{Eqn: Spectrum of the Morlet wavelet} 
\end{eqnarray}

From Eq.~(\ref{Eqn: Spectrum of the Morlet wavelet}) it is clear
that the admissibility condition \( \hat{\psi }_{\textrm{M}}(0;\sigma )=0 \)
implies that \( \kappa (\sigma )=e^{-\frac{1}{2}\sigma ^{2}}. \)

Many previous applications of the Morlet wavelet have been concerned
with signals containing slowly varying frequency and amplitude components
for which large values of \( \sigma  \) (\( \geq 5 \)) are applicable
and \( \kappa (\sigma ) \) (\( \leq 10^{-6} \)) is negligible~\cite{Goupillaud1984}.

However, we are interested in applying this type of analysis to signals
which contain rapidly varying frequencies and amplitudes. In this
case, the ability to use small values of \( \sigma  \) becomes important
as we wish to maximize the temporal resolution by minimizing \( \sigma  \)
whilst still being able to separate the various frequency components
in the signal and, consequently, \( \kappa (\sigma ) \) is no longer
negligible.

\begin{figure}
{\centering \resizebox*{0.4\textwidth}{!}{\includegraphics{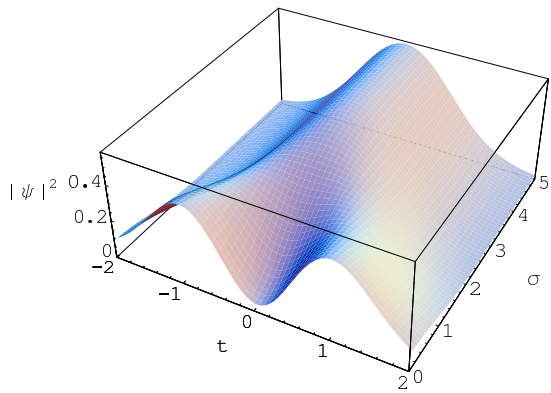}} \par}

\caption{\label{Fig: Envelope of the Morlet wavelet}Envelope \protect\( |\psi _{\textrm{M}}|^{2}\protect \)
of the Morlet wavelet \protect\( \psi _{\textrm{M}}(t;\sigma )\protect \)
(Eq.~\ref{Eqn: Morlet wavelet}) showing the unwanted transition
from mono-modal to bimodal behaviour at small \protect\( \sigma \protect \)
(\protect\( <1.79785\protect \)).}
\end{figure}

\begin{figure}
{\centering \resizebox*{0.4\textwidth}{!}{\includegraphics{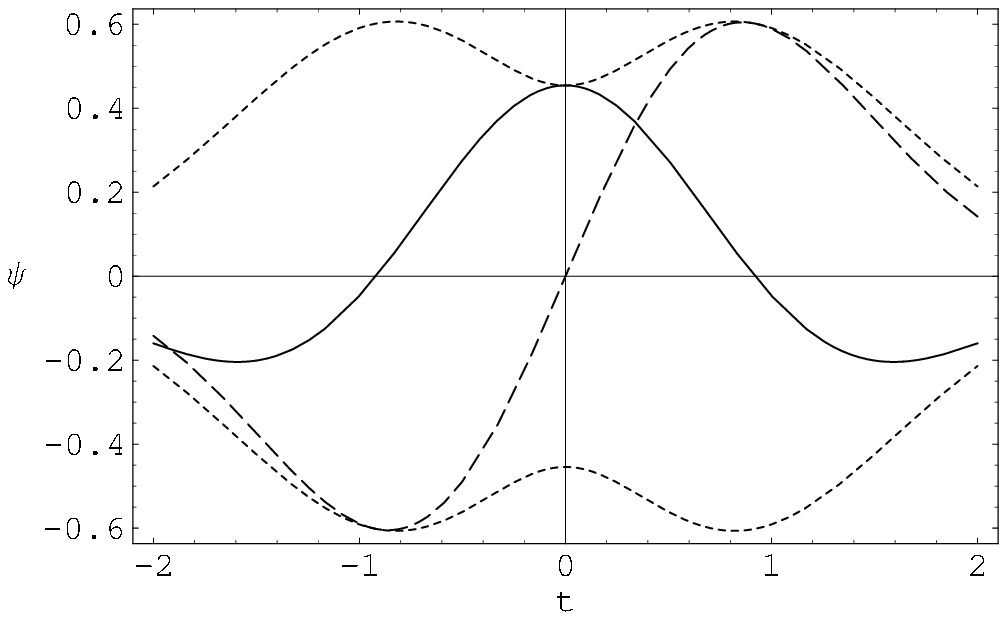}} \par}

\caption{\label{Fig: Morlet wavelet sigma=3D1}Morlet wavelet \protect\( \psi _{\textrm{M}}(t;\sigma )\protect \)
(Eq.~\ref{Eqn: Morlet wavelet}) for \protect\( \sigma =1\protect \):
real part, solid line; imaginary part, long dashed line; and envelope
\protect\( \pm |\psi |\protect \), short dashed lines.}
\end{figure}

Although the Morlet wavelet is admissible at small \( \sigma  \),
the temporal localization is unsatisfactory (see Fig.~\ref{Fig: Envelope of the Morlet wavelet});
namely, \( |\psi _{\textrm{M}}|^{2} \) undergoes a transition from
mono-modality to bimodality (a single ridge at large \( \sigma  \)
splits into two symmetric ridges for small \( \sigma  \)). The wavelet
transform of a signal performed using a wavelet which has a bimodal
envelope results in the signal being localized about two different
positions (see Fig.~\ref{Fig: Morlet wavelet sigma=3D1}). This produces
unwanted artefacts in the resulting instantaneous frequency and amplitude
measurements (shown later in Figs.~\ref{Fig: Example function WT period}
and \ref{Fig: Example function WT amplitude}).

Therefore we remedy this drawback by modifying the Morlet wavelet
to produce a new wavelet, \( \psi (t;\sigma ) \), such that \( |\psi |^{2} \)
has a single, global maximum for all \( \sigma  \). For the new wavelet
we choose to replace the single, normalization constant \( c_{\textrm{M}}(\sigma ) \)
in the Morlet wavelet with two new parameters \( p(\sigma ) \) and
\( q(\sigma ) \) determined by two conditions: (i) total normalization
of the wavelet to unity, and (ii) equal contributions to the normalization
from the real and imaginary parts. The new wavelet has the following
functional form:\begin{eqnarray}
\psi (t;\sigma ) & =\, \pi ^{-\frac{1}{4}}e^{-\frac{1}{2}t^{2}}\! \! \!  & \left[ p(\sigma )\left( \cos (\sigma t)-\kappa (\sigma )\right) \right. \nonumber \\
 &  & \left. +\, i\, q(\sigma )\sin (\sigma t)\right] ,\label{Eqn: Harrop wavelet} 
\end{eqnarray}
where \( p(\sigma ) \) and \( q(\sigma ) \) are given by:\begin{subequations}\begin{eqnarray}
p(\sigma ) & = & \left( 1-e^{-\sigma ^{2}}\right) ^{-\frac{1}{2}},\label{Eqn: Harrop wavelet constant 1} \\
q(\sigma ) & = & \left( 1+3e^{-\sigma ^{2}}-4e^{-\frac{3}{4}\sigma ^{2}}\right) ^{-\frac{1}{2}}.\label{Eqn: Harrop wavelet constant 2} 
\end{eqnarray}
\end{subequations}

The Fourier transform of this wavelet is:\begin{eqnarray}
\hat{\psi }(\omega ;\sigma ) & = & \frac{1}{2}e^{-\frac{1}{2}\left( \sigma +\omega \right) ^{2}}\left( e^{\sigma \omega }-1\right) \nonumber \\
 &  & \times \left[ \left( e^{\sigma \omega }-1\right) p(\sigma )+\left( e^{\sigma \omega }+1\right) q(\sigma )\right] .\label{Eqn: Harrop wavelet spectrum} 
\end{eqnarray}

\begin{figure}
{\centering \resizebox*{0.4\textwidth}{!}{\includegraphics{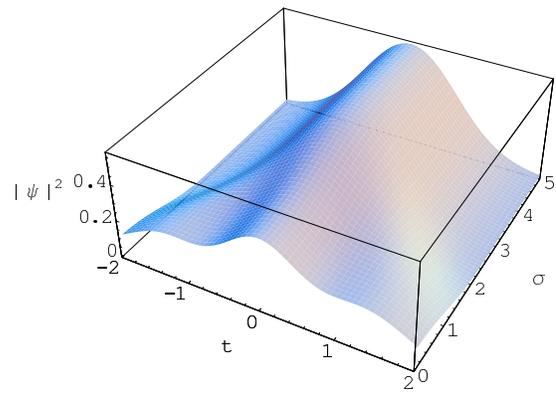}} \par}

\caption{\label{Fig: Envelope of the Harrop wavelet}Envelope \protect\( |\psi |^{2}\protect \)
of the new wavelet (Eq.~\ref{Eqn: Harrop wavelet}).}
\end{figure}

\begin{figure}
{\centering \resizebox*{0.4\textwidth}{!}{\includegraphics{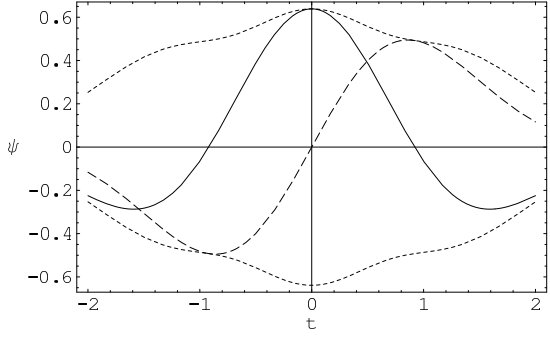}} \par}

\caption{\label{Fig: Harrop wavelet sigma=3D1}New wavelet \protect\( \psi (t;\sigma )\protect \)
(Eq.~\ref{Eqn: Harrop wavelet}) for \protect\( \sigma =1\protect \):
real part, solid line; imaginary part, long dashed line; and envelope
\protect\( \pm |\psi |\protect \), short dashed lines.}
\end{figure}

It is worthwhile noting that the real part of this new wavelet recovers
the functional form of the Mexican Hat wavelet in the limit \( \sigma \rightarrow 0 \):\[
\textrm{Re}\left[ \psi (t;0)\right] =\sqrt{\frac{2}{3}}\pi ^{-\frac{1}{4}}e^{-\frac{1}{2}t^{2}}(t^{2}-1).\]
Thus the new wavelet allows a complete transition from very high temporal
localization, \( \sigma \rightarrow 0 \) (the Mexican Hat wavelet),
to maximum frequency localization, \( \sigma \rightarrow \infty  \)
(plane wave). Even in the limit of minimal \( \sigma  \), \( |\psi |^{2} \)
remains mono-modal (see Figs.~\ref{Fig: Envelope of the Harrop wavelet}
and \ref{Fig: Harrop wavelet sigma=3D1}). Thus we have improved upon
the temporal localisation of the Morlet wavelet.

We have also checked that, using the new wavelet, the original signal
can be recovered by the resolution of the identity operator.

\section{Analysis\label{Sec: Analysis}}

\subsection{Instantaneous Frequency and Amplitude\label{Sec: Instantaneous Frequency and Amplitude}}

In this section we demonstrate how the new wavelet may be used to
extract instantaneous frequencies and amplitudes from a signal via
the CWT. The following analysis is based upon the stationary-phase
approach of Delprat et al.~\cite{Delprat1991} but is specialized
to the new wavelet.

The wavelet transform \( F(a,b) \) at a given scale \( a \) and
translation \( b \) is given by the integral (Eq.~\ref{Eqn: Integral expression for the wavelet transform})
of a rapidly oscillating integrand. This integral may be rewritten
in the form:\begin{eqnarray}
F(a,b) & \equiv  & \frac{1}{2}\int _{-\infty }^{\infty }e^{i\Phi (t;a,b)+\ln A(t;a,b)}\, dt,\label{Eqn: Wavelet transform Complex Integrand} 
\end{eqnarray}
where:\begin{subequations}\begin{eqnarray}
A(t;a,b) & = & A_{f}(t)\, A_{\xi }\! \left( t;a,b\right) \label{Eqn: Amplitude of stationary phase integrand} \\
\Phi (t;a,b) & = & \phi _{f}(t)-\phi _{\xi }\! \left( t;a,b\right) \label{Eqn: Phase of stationary phase integrand} 
\end{eqnarray}
\end{subequations}with \( f(t)=\textrm{Re}[A_{f}(t)e^{i\phi _{f}(t)}] \)
and \( \xi (t;a,b)=A_{\xi }(t;a,b)e^{i\phi _{\xi }(t;a,b)} \).

In order to take the integral in the stationary-phase approximation,
we first approximate \( A_{\xi } \) by a Gaussian. From Eq.~\ref{Eqn: Dilation and translation of the mother wavelet}
we have \( A_{\xi }^{2}(t;a,b)=a^{-2}A_{\psi }^{2}((t-b)/a) \), where
\( A_{\psi }^{2} \) is taken to be a normalized Gaussian whose variance
\( \sigma _{\psi }(\sigma ) \) is equal to the variance of \( |\psi |^{2} \),
giving:\begin{equation}
\label{Eqn: Approximate psi envelope}
A^{2}_{\xi }\! \left( t;a,b\right) \simeq |a|^{-2}\frac{1}{\sqrt{2\pi }\sigma _{\psi }}\exp \! \left[ -\frac{1}{2\sigma _{\psi }^{2}}\left( \frac{t-b}{a}\right) ^{2}\right] ,
\end{equation}
where the variance \( \sigma _{\psi }(\sigma ) \) can be found analytically:\begin{eqnarray}
\sigma _{\psi }^{2}(\sigma )= &  & \frac{1}{4}\sqrt{\pi }\left[ q^{2}\left( \left( 2\sigma ^{2}-1\right) e^{-\sigma ^{2}}+1\right) \right. \label{Eqn: sigma_psi} \\
 &  & +\left. p^{2}\left( \left( 3-2\sigma ^{2}\right) e^{-\sigma ^{2}}-2e^{-\frac{3}{4}\sigma ^{2}}\left( 2-\sigma ^{2}\right) \right) \right] .\nonumber 
\end{eqnarray}

\begin{figure*}
{\centering a)\resizebox*{0.4\textwidth}{!}{\includegraphics{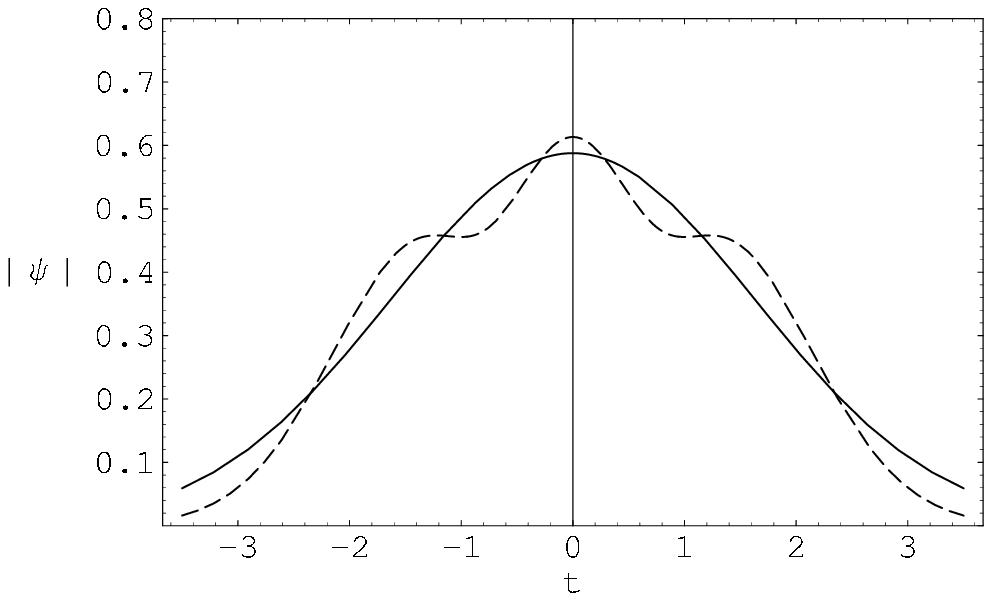}} 
b)\resizebox*{0.4\textwidth}{!}{\includegraphics{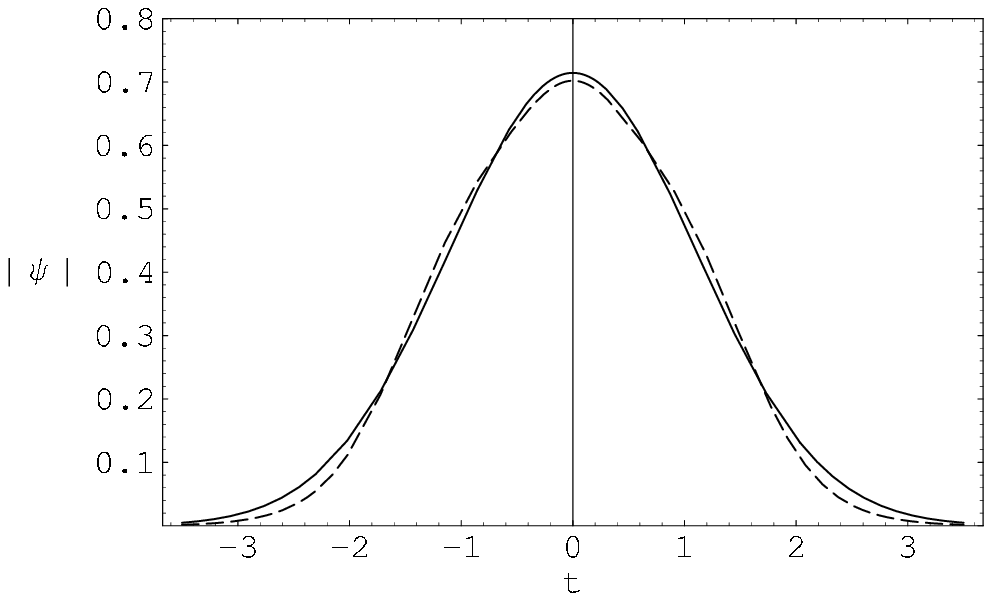}} \par}

\caption{\label{Fig: Harrop wavelet Gaussian approximations}Gaussian approximations
(solid lines, Eq.~\ref{Eqn: Approximate psi envelope}) to the true
envelope \protect\( |\psi (t;\sigma )|^{2}\protect \) of the mother
wavelet function (dashed lines, Eq.~\ref{Eqn: Harrop wavelet}) for:
a) \protect\( \lim \sigma \rightarrow 0\protect \); b) \protect\( \sigma =2\protect \).}
\end{figure*}

The approximate envelope, \( A_{\psi }^{2} \), tends to the true
envelope, \( |\psi |^{2} \) (see Fig.~\ref{Fig: Harrop wavelet Gaussian approximations}).

We assume (without loss of generality~\cite{Delprat1991}) that there
is a single point of stationary phase for the integrand in Eq.~(\ref{Eqn: Wavelet transform Complex Integrand})
at \( t=t_{s}(a,b) \). Under the conventional asymptotic approximation:\begin{equation}
\label{Eqn: Asymptotic signal}
\left| \frac{\partial \phi _{f}}{\partial t}\right| \gg \left| \frac{1}{A}\frac{\partial A_{f}}{\partial t}\right| ,
\end{equation}
we expand \( \Phi (t) \) around the stationary point \( t_{s} \)
assuming \( \Phi ''(t_{s})\neq 0 \) and substitute the approximate
expression for \( A_{\xi } \) from Eq.~(\ref{Eqn: Approximate psi envelope})
into the integral, which can then be taken. This gives an approximate
expression for the squared modulus of the CWT using the new wavelet:
\begin{eqnarray}
\left| F(a,b)\right| ^{2} & \simeq  & \sqrt{\frac{\pi }{2}}\sigma _{\psi }A_{f}^{2}(t_{s})\left( 1+4a^{4}\sigma _{\psi }^{4}\Phi ''(t_{s})^{2}\right) ^{-\frac{1}{2}}\nonumber \\
 &  & \times \exp \! \left[ -\frac{a^{2}\sigma _{\psi }^{2}\Phi ''(t_{s})^{2}(t_{s}-b)^{2}}{1+4a^{4}\sigma _{\psi }^{4}\Phi ''(t_{s})^{2}}\right] .\label{Eqn: Accurate WT modulus} 
\end{eqnarray}

Further, assuming the frequency of the mother wavelet to be constant
(\( \phi _{\psi }''(t)=0 \)) and the frequency variation of the signal
to be slow in the region of interest (i.e.~\( |\Phi ''(t_{s})|a^{2}\sigma _{\psi }^{2}\ll 1 \))
then:\begin{equation}
\label{Eqn: Slow frequency variation form of the WT}
\left| F(a,b)\right| ^{2}\simeq \sqrt{\frac{\pi }{2}}\sigma _{\psi }A_{f}^{2}(t_{s})e^{-a^{2}\sigma _{\psi }^{2}\Phi ''(t_{s})^{2}(t_{s}-b)^{2}}.
\end{equation}

For a monochromatic signal (i.e.~a signal which contains only a single
frequency at any given position), there is a scale \( a_{r}(b) \)
at any given \( b \) which corresponds to a basis wavelet centred
at \( b \) whose frequency \( \phi _{\xi }'(b;a_{r}(b),b) \) is
equal to the local frequency of \( f(t) \). The scale \( a_{r}(b) \)
of this wavelet identifies the instantaneous frequency of the signal
and may be found as the solution of the equation \( \Phi '(b;a_{r}(b),b)=0 \).
From the definition of the points of stationary phase (\( \Phi '(t_{s}(a,b);a,b)=0 \)),
this corresponds to \( t_{s}(a_{r}(b),b)=b \), an alternative equation
which can be used to find \( a_{r}(b) \). With the choice of normalization
used in Eq.~(\ref{Eqn: Dilation and translation of the mother wavelet}),
it is clear that these points maximize the expression for the squared
modulus of the CWT with respect to \( a \) as obtained by the stationary-phase
approximation, Eq.~(\ref{Eqn: Slow frequency variation form of the WT}).

As the CWT is a linear operation, superimposed frequency components
are manifested as different scales \( a_{r}^{(i)}(b) \) which locally
maximize \( |F| \) (assuming sufficiently large \( \sigma  \) to
resolve the peaks). The curves formed by the points \( a_{r}^{(i)}(b),b \)
are known as the {}``ridges'' of the transform~\cite{Delprat1991}.
The trajectory of each ridge can be used to extract the amplitude
and frequency modulations of the corresponding signal components.

An approximate expression for the instantaneous amplitude, \( A_{f}(t) \),
of a signal component can be obtained by rewriting the stationary-phase
approximation to the squared modulus of the WT (Eq.~\ref{Eqn: Slow frequency variation form of the WT})
on the ridge, \( |F(a_{r}^{(i)}(b),b)|^{2}\simeq \sqrt{\pi /2}\, \sigma _{\psi }A_{f}^{(i)}(b)^{2} \),
in terms of \( A_{f}(t) \):\begin{equation}
\label{Eqn: Instantaneous amplitude}
A_{f}^{(i)}(t)\simeq \left( \frac{1}{2}\pi \sigma _{\psi }^{2}\right) ^{-\frac{1}{4}}\left| F(a_{r}^{(i)}(t),t)\right| .
\end{equation}

There are two different well-known approximations to the instantaneous
frequency \( \phi _{f}'(t)/2\pi  \). As each has relative merits,
we consider both.

The simplest approximation to the instantaneous frequency is the rate
of change of the phase of the CWT with respect to \( b \), evaluated
at \( a_{r}(b),b \):\begin{equation}
\label{Eqn: Instantaneous frequency from arg}
\nu _{f}^{(i)}(t)\simeq \frac{1}{2\pi }\left| \left[ \frac{\partial }{\partial b}\textrm{Arg}[F(a_{r}^{(i)}(b),b)]\right] _{b=t}\right| .
\end{equation}
The derivation for this expression using the new wavelet is identical
to that of the Morlet wavelet given by Delprat et al.~\cite{Delprat1991}.

The other approximation to the instantaneous frequency uses the equality
of the frequency of the signal and of the wavelet on a ridge to create
an expression for the frequency of the signal as a function of the
scale \( a_{r}(b) \) on the ridge and the frequency of the mother
wavelet, \( \omega _{\psi }(\sigma ) \):\begin{equation}
\label{Eqn: Instantaneous frequency from mod}
\nu _{f}^{(i)}(t)\simeq \frac{\omega _{\psi }}{2\pi }\left| a_{r}^{(i)}(t)\right| ^{-1}.
\end{equation}

\begin{figure}
{\centering \resizebox*{0.4\textwidth}{!}{\includegraphics{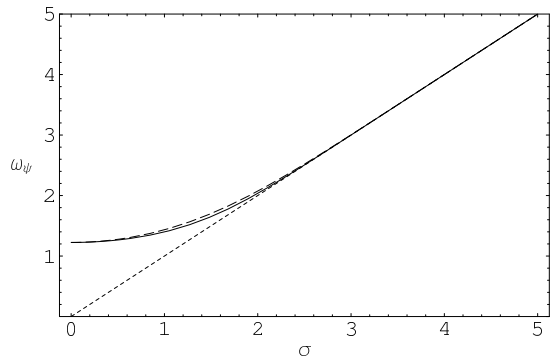}} \par}

\caption{\label{Fig: Harrop wavelet mean and mode frequency}Approximations
to the frequency \protect\( \omega _{\psi }\protect \) of the mother
wavelet function (Eq.~\ref{Eqn: Harrop wavelet}): Analytic mean
of \protect\( |\hat{\psi }|^{2}\protect \) (solid line, Eq.~\ref{Eqn: Harrop wavelet mean frequency}),
numerical mode of \protect\( |\hat{\psi }|^{2}\protect \) (long dashed
line) and the asymptotic approximation \protect\( \omega _{\psi }=\sigma \protect \)
valid in the limit \protect\( \sigma \rightarrow \infty \protect \)
(short dashed line).}
\end{figure}

Conventionally, \( \omega _{\psi } \) is taken to be the underlying
modulating frequency, \( \sigma  \), of the mother wavelet function.
However, this is a poor approximation at small \( \sigma  \). Therefore,
the obvious definition of \( \omega _{\psi } \) is the modal average
(position of the highest peak) in the Fourier power spectrum \( |\hat{\psi }|^{2} \).
Unfortunately, this expression for \( \omega _{\psi } \) cannot be
found analytically. However, even at small \( \sigma  \), the spectrum
\( |\hat{\psi }|^{2} \) is nearly symmetric about the main peak (see
Fig.~\ref{Fig: Harrop wavelet spectrum at sigma=3D0}). Therefore,
the mean average is always a good approximation to the modal average
(see Fig.~\ref{Fig: Harrop wavelet mean and mode frequency}) and,
unlike the mode, the mean can be found analytically:
\begin{figure}
{\centering \resizebox*{0.4\textwidth}{!}{\includegraphics{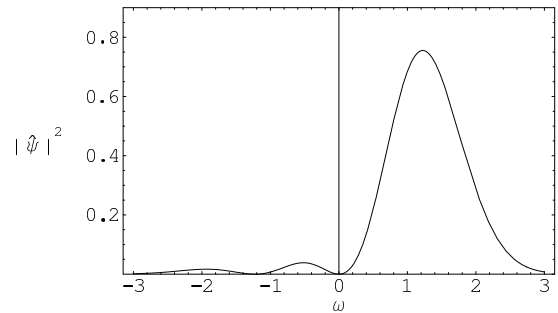}} \par}

\caption{\label{Fig: Harrop wavelet spectrum at sigma=3D0}Fourier power spectrum
\protect\( |\hat{\psi }(\omega ,\sigma )|^{2}\protect \) of the mother
wavelet function (c.f.~Eq.~\ref{Eqn: Harrop wavelet spectrum})
in the limit \protect\( \sigma \rightarrow 0\protect \).}
\end{figure}

\begin{eqnarray}
\omega _{\psi }(\sigma ) & = & \sqrt{\pi }\sigma \, p(\sigma )\, q(\sigma )\left( 1-e^{-\frac{3}{4}\sigma ^{2}}\right) .\label{Eqn: Harrop wavelet mean frequency} 
\end{eqnarray}

Using this expression for \( \omega _{\psi } \) in conjunction with
the relationship between scale and frequency in Eq.~(\ref{Eqn: Instantaneous frequency from mod}),
a CWT may be plotted as a function \( |F(\omega _{\psi }/2\pi \nu ,t)| \)
of time and frequency.

Delprat et al.~\cite{Delprat1991} proposed that the phase-based
instantaneous frequency, Eq.~(\ref{Eqn: Instantaneous frequency from arg}),
is more accurate than the modulus-based measurement, Eq.~(\ref{Eqn: Instantaneous frequency from mod}),
and suggested an iterative algorithm for extracting signal components.
Carmona et al.~have since shown that the modulus-based measurement
is extremely resiliant to noise~\cite{Carmona1997} and have suggested
numerous methods for extracting signal components using this approach~\cite{Carmona1999}.

Thus the instantaneous frequencies and amplitudes of components in
a signal may be found from the CWT at the points where \( |F(a,b)| \)
is locally maximized with respect to \( a \). These maxima can be
identified numerically from a set of samples of \( F(a,b) \) generated
by discrete approximation to the integral expression for the CWT,
Eq.~(\ref{Eqn: Integral expression for the wavelet transform}).
Once found, the maxima may be interpreted using the approximate analytic
results given above.

\subsection{Example Function\label{Sec: Example Function}}

The method described in the previous section is most easily clarified
by the following examples. First, we choose to apply the method to
the simple, variable-frequency function (see Fig.~\ref{Fig: Example function}):
\begin{figure}
{\centering \resizebox*{0.4\textwidth}{!}{\includegraphics{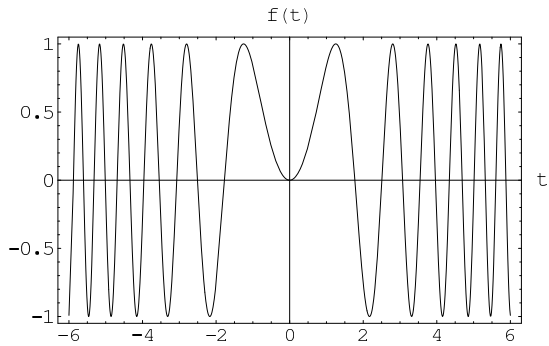}} \par}

\caption{\label{Fig: Example function}Variable-frequency function \protect\( f(t)\protect \)
(Eq.~\ref{Eqn: Example function}).}
\end{figure}

\begin{equation}
\label{Eqn: Example function}
f(t)=\sin \! \left( t^{2}\right) .
\end{equation}

The FT \( \hat{f}(\omega )=\frac{1}{2}\left( \cos \! \left( \frac{1}{4}\omega ^{2}\right) -\sin \! \left( \frac{1}{4}\omega ^{2}\right) \right)  \)
conveys little useful information about the original function.

\begin{figure}
{\centering \resizebox*{0.4\textwidth}{!}{\includegraphics{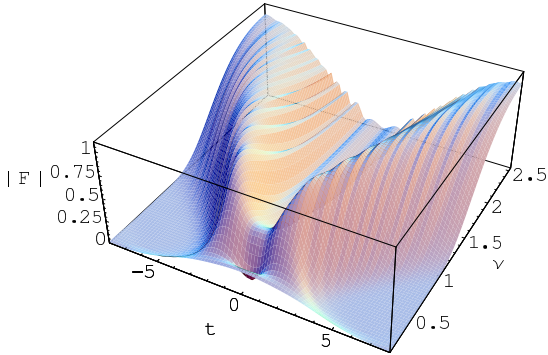}} \par}

\caption{\label{Fig: Example function WT}Modulus of the CWT as a function
of frequency \protect\( F(\omega _{\psi }/2\pi \nu ,t)\protect \)
of the function \protect\( f(t)\protect \) (Eq.~\ref{Eqn: Example function})
using the new mother wavelet function \protect\( \psi (t;\sigma )\protect \)
(Eq.~\ref{Eqn: Harrop wavelet}) with \protect\( \sigma =2\protect \).}
\end{figure}

However, the modulus of the WT does convey useful information, particularly
when plotted as a function of frequency instead of scale (see Fig.~\ref{Fig: Example function WT})
as this highlights the linearly changing local frequency of \( f(t) \)
(given by \( \nu _{f}=|\partial /\partial t\, \phi _{f}|/2\pi  \))
as a function of \( t \).

\begin{figure*}
{\centering a)\resizebox*{0.4\textwidth}{!}{\includegraphics{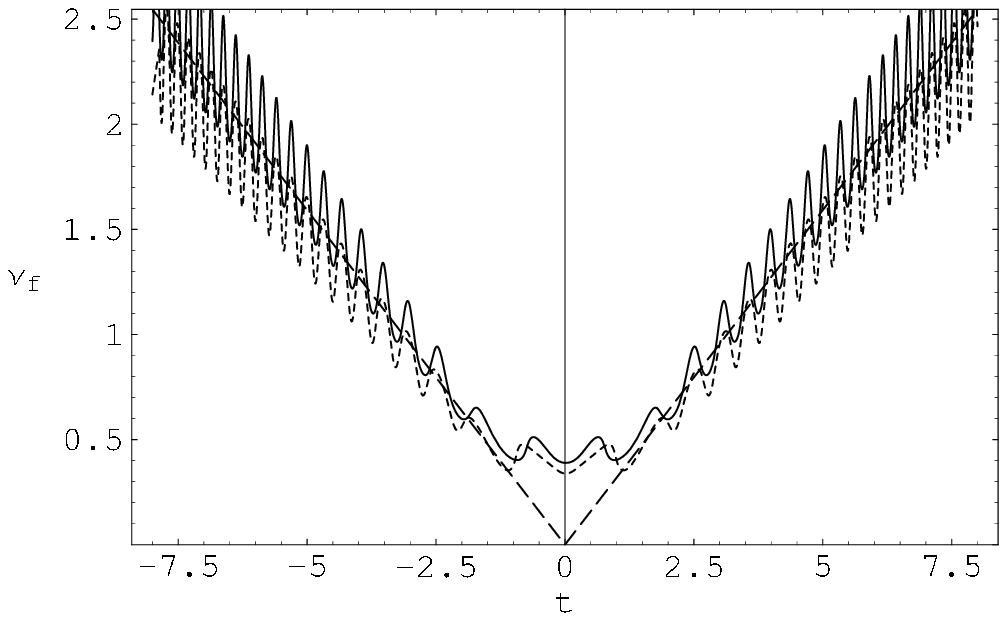}} 
b)\resizebox*{0.4\textwidth}{!}{\includegraphics{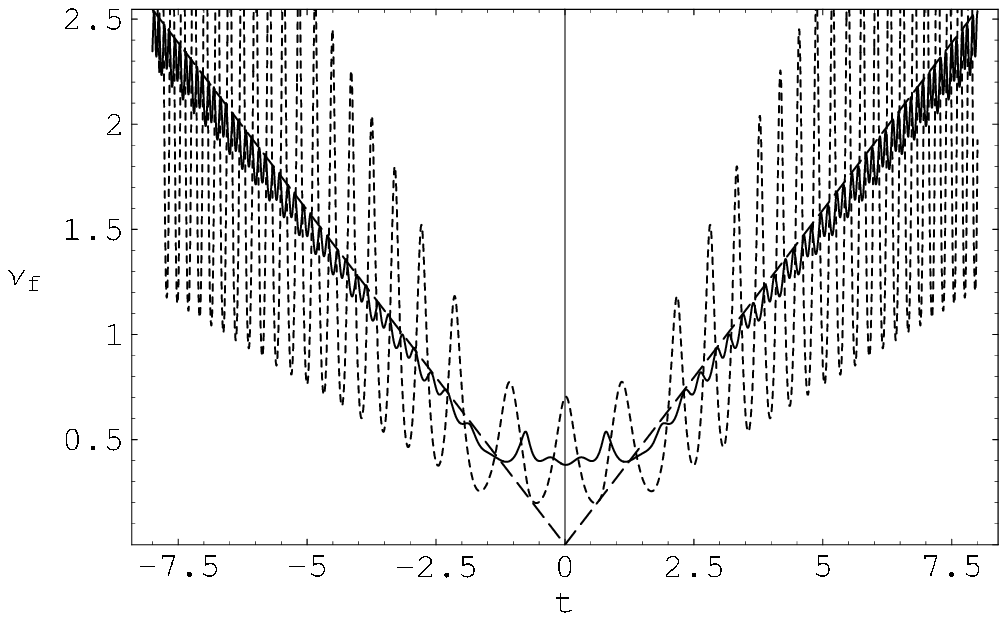}} \par}

\caption{\label{Fig: Example function WT period}Instantaneous frequencies
\protect\( \nu _{f}(t)\protect \) using the new wavelet (solid lines),
Morlet wavelet (short dashed lines) and expected value \protect\( \nu (t)=|t|/\pi \protect \)
for \protect\( |t|\gg 0\protect \) (long dashed lines) of the example
function \protect\( f(t)\protect \) (Eq.~\ref{Eqn: Example function})
with \protect\( \sigma =1\protect \). Extracted using: a) maximal
\protect\( |F|\protect \), Eq.~(\ref{Eqn: Instantaneous frequency from mod});
b) \protect\( \partial \textrm{Arg}[F]/\partial b\protect \), Eq.~(\ref{Eqn: Instantaneous frequency from arg}).}
\end{figure*}

\begin{figure}
{\centering \resizebox*{0.4\textwidth}{!}{\includegraphics{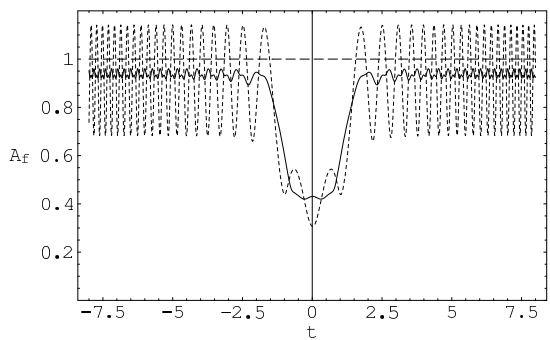}} \par}

\caption{\label{Fig: Example function WT amplitude}Instantaneous amplitude
\protect\( A_{f}(t)\protect \) (Eq.~\ref{Eqn: Instantaneous amplitude})
of \protect\( f(t)\protect \) (Eq.~\ref{Eqn: Example function})
with \protect\( \sigma =1\protect \). Instantaneous amplitude \protect\( A_{f}(t)\protect \),
solid line; expected amplitude of \protect\( 1.0\protect \) for \protect\( |t|\gg 0\protect \),
dashed line.}
\end{figure}

The CWT of \( f(t) \) contains a single, `V' shaped ridge at \( a_{r}(b) \).
This ridge reflects both the frequency modulation of \( f(t) \) (see
Fig.~\ref{Fig: Example function WT period}) and the amplitude modulation
(see Fig.~\ref{Fig: Example function WT amplitude}). In all cases,
the results show fluctuations linked with the phase \( \phi _{f} \)
of the signal. However, compared to the Morlet wavelet, the new wavelet
produces much smaller fluctuations in all results.

\subsection{Reduced Radial Distribution Function\label{Sec: Reduced Radial Distribution Function}}

We now apply the method described in Sec.~3.1 to a function of practical
interest. We choose to study the RRDF \( d(r) \) of a model glass
structure.

The RRDF analyzed in this paper is taken from a structural model of
the icosahedral (IC) glass~\cite{Dzugutov1993} created in a classical
molecular-dynamics simulation~\cite{Simdyankin2000}. We calculate
the transform as detailed in Sec.~2 and perform the analysis as discussed
in Sec.~3 in order to study the components of \( d(r) \). The function
\( d(r) \) is considered to be zero outside the range \( 0<r<L/2 \),
where \( L/2\simeq 25 \) is half the side of the cubic simulation
super-cell which contains \( 108,000 \) atoms.

\begin{figure*}
{\centering a)\resizebox*{0.4\textwidth}{!}{\includegraphics{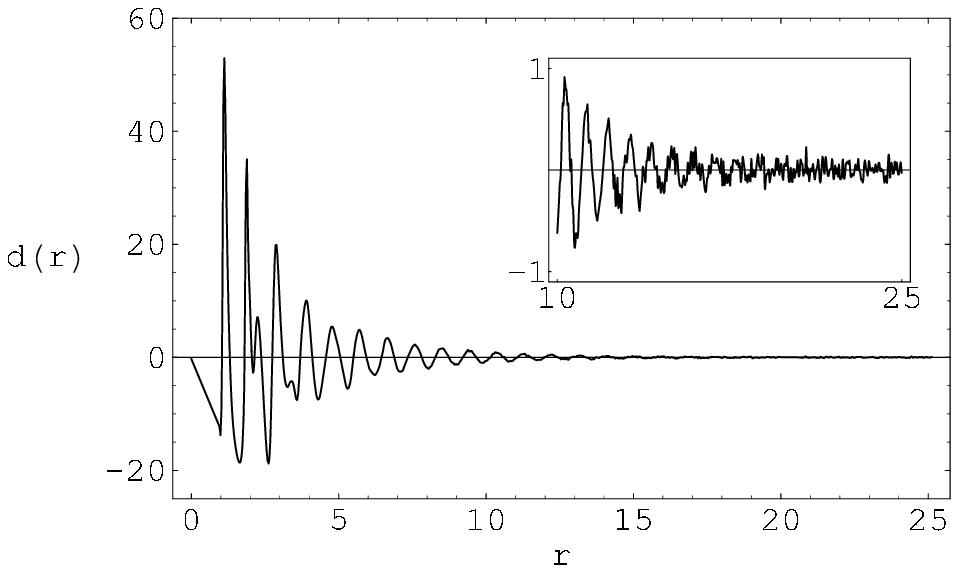}} 
b)\resizebox*{0.4\textwidth}{!}{\includegraphics{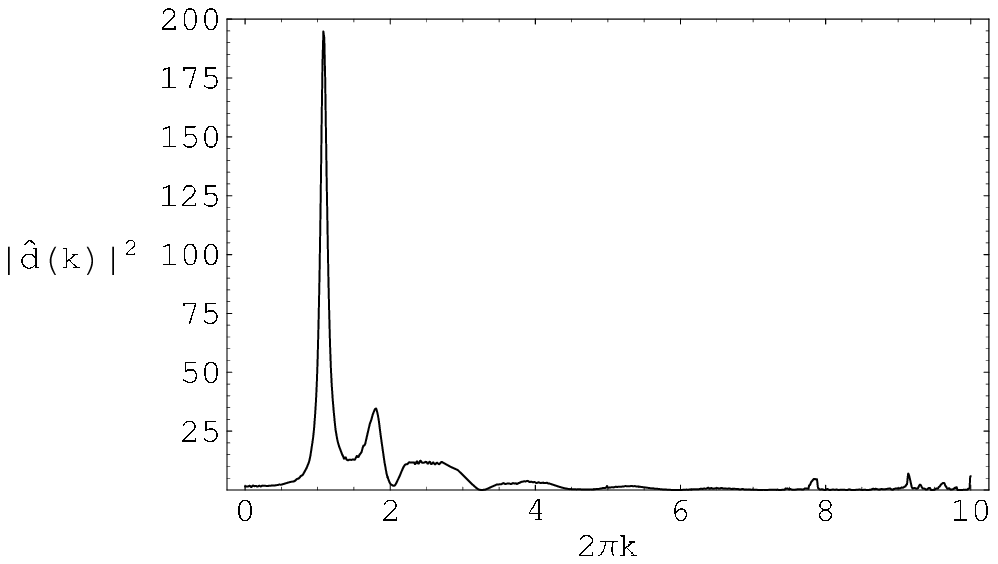}} \par}

\caption{\label{Fig: RRDF}a) RRDF of the IC glass (inset showing a magnification
of the damped density fluctuations); b) its Fourier power spectrum
\protect\( |\hat{d}(k)|^{2}\protect \).}
\end{figure*}

The RRDF, \( d(r) \), is defined in terms of the atomic density \( \rho (r) \)
as:\begin{equation}
\label{Eqn: Definition of d(r)}
d(r)=4\pi r\left( \rho (r)-\rho _{0}\right) ,
\end{equation}
where \( \rho _{0} \) is the average atomic density~\cite{Elliott1990}.
This is shown in Fig.~\ref{Fig: RRDF}a for the IC glass. Reduced
Lennard-Jones units (r.u.) are used for length with the mean nearest-neighbour
separation being \( 1.15\pm 0.05 \)r.u.. The damped extended-range
density fluctuations are clearly visible, extending beyond \( 10 \)r.u.~(see
the inset in Fig.~\ref{Fig: RRDF}a).

From the Fourier power spectrum of \( d(r) \) (shown in Fig.~\ref{Fig: RRDF}b),
it is clear that \( d(r) \) contains many components with different
frequencies. The highest peak in \( |\hat{d}|^{2} \) occurs at the
frequency \( \nu _{d}=1.08\pm 0.01 \). This peak has non-zero width
implying that the real-space fluctuation in \( d(r) \) corresponding
to this peak has a spatially varying amplitude but we cannot deduce
a functional form from this alone.

\begin{figure*}
{\centering a)\resizebox*{0.4\textwidth}{!}{\includegraphics{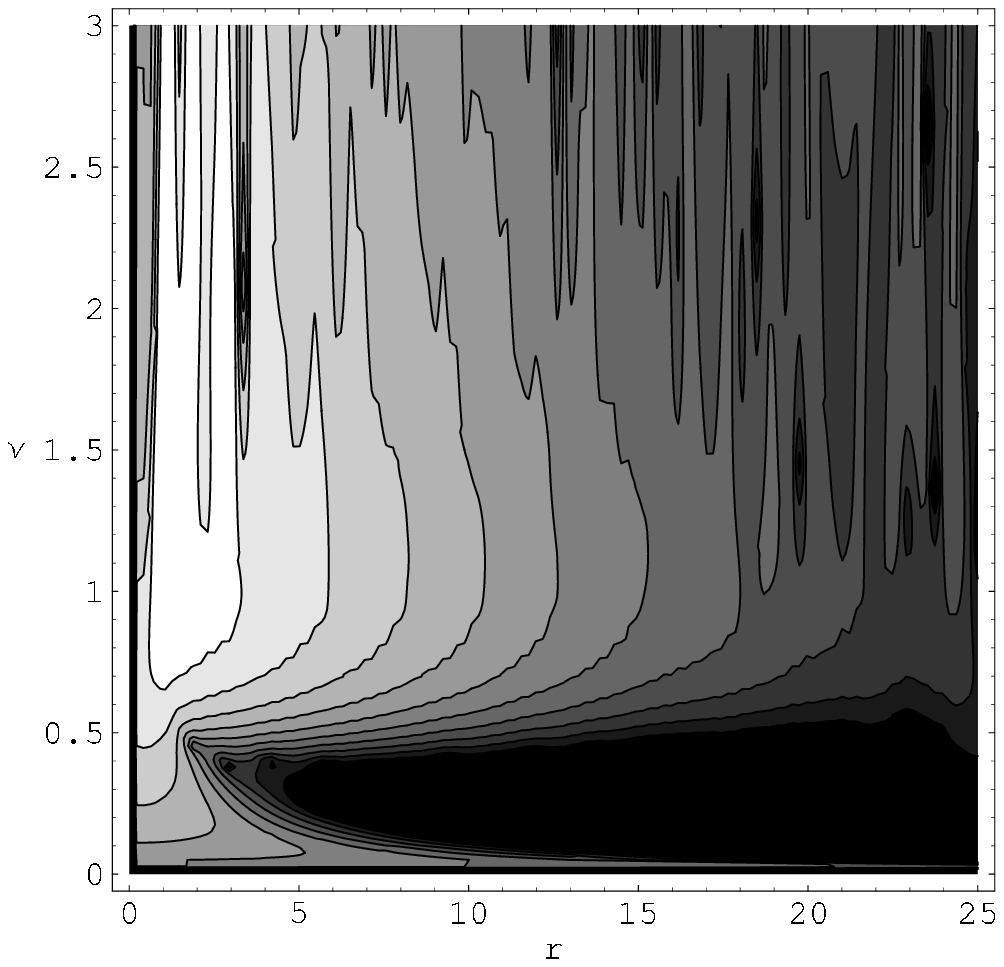}} 
b)\resizebox*{0.4\textwidth}{!}{\includegraphics{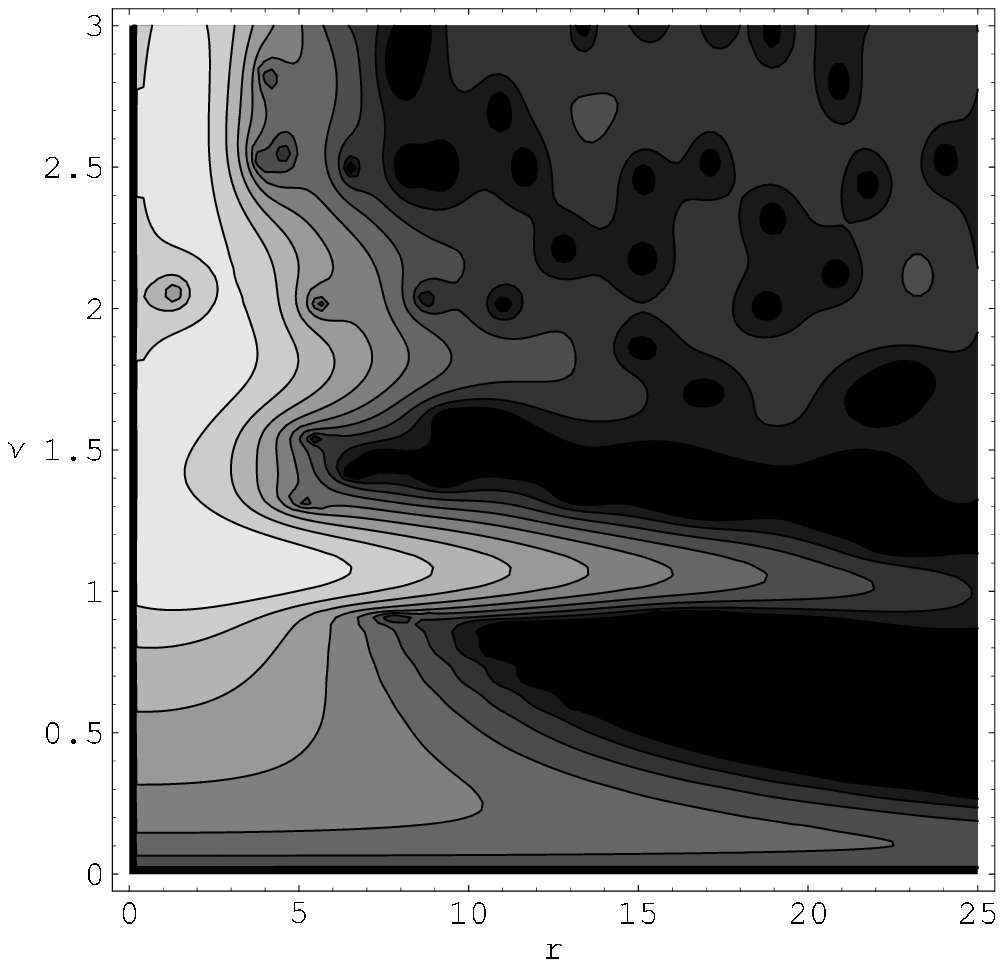}} \par}

\caption{\label{Fig: RRDF WT}Modulus \protect\( |F_{d}(\omega _{\psi }/2\pi \nu ,r)|\protect \)
of the CWT of the RRDF \protect\( d(r)\protect \) plotted as a function
of frequency \protect\( \nu \protect \) (see Fig.~\ref{Fig: RRDF})
using the new wavelet (Eq.~\ref{Eqn: Harrop wavelet}) for: a) \protect\( \sigma =2\protect \);
b) \protect\( \sigma =15\protect \).}
\end{figure*}

Plotting the modulus \( |F_{d}(a,b)| \) of the CWT using different
envelope widths, shown as a function of \( r \) and \( \nu  \) (\( \equiv 2\pi k \))
in Fig.~\ref{Fig: RRDF WT}, allows \( d(r) \) to be examined in
the time-frequency plane. Using small \( \sigma  \) results in high
spatial resolution but poor frequency resolution and the ridges are
smeared together (see Fig.~\ref{Fig: RRDF WT}a). Larger values of
\( \sigma  \) separate the ridges at the cost of decreasing the spatial
resolution (see Fig.~\ref{Fig: RRDF WT}b). Unlike the example function
from the previous section, \( d(r) \) contains several components
with different frequencies which, particularly when using large \( \sigma  \),
manifest themselves as separate ridges in the WT. In this paper we
consider only the prominent ridge at \( \nu \simeq \nu _{d} \) but
the same analysis can be applied to the ridges seen at other frequencies.

The ridge along \( \nu \simeq \nu _{d} \) shows that the prominent
frequency component identified from the Fourier power spectrum of
\( d(r) \) is particularly strong around \( r=0 \) but decays away
with greater \( r \). This trajectory of the ridge can then be used
to extract the instantaneous frequencies and amplitudes of this component
in \( d(r) \).

\begin{figure}
{\centering \resizebox*{0.4\textwidth}{!}{\includegraphics{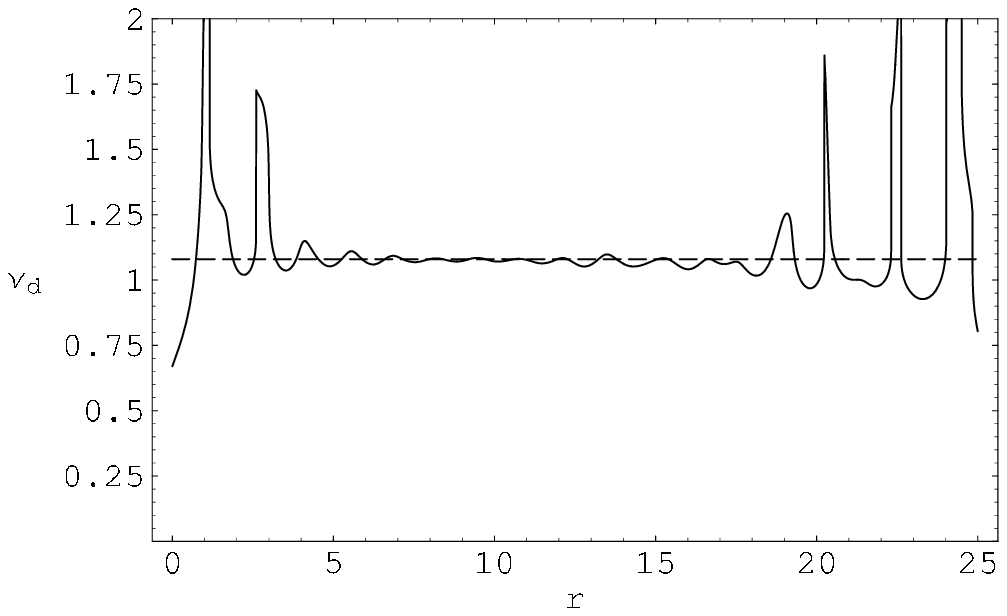}} \par}

\caption{\label{Fig: RRDF Instantaneous frequency from mod}Instantaneous
frequency \protect\( \nu _{d}(r)\protect \) of the largest component
of the RRDF \protect\( d(r)\protect \) (see Fig.~\ref{Fig: RRDF}b).
Solid line is \protect\( \nu _{d}\protect \) extracted using Eq.~(\ref{Eqn: Instantaneous frequency from mod})
with \protect\( \sigma =3\protect \), dashed line is the best-fit
constant frequency \protect\( \nu _{d}=1.08\protect \) over the range
\protect\( 5<r<18\protect \).}
\end{figure}

The instantaneous frequency found using \( \sigma =3 \) (see Fig.~\ref{Fig: RRDF Instantaneous frequency from mod})
remains constant over a large range of \( r \). As expected, the
scale at which this ridge occurs in the CWT of \( d(r) \) corresponds
to the position of the prominent peak in the Fourier power spectrum
of \( d(r) \).

\begin{figure}
{\centering \resizebox*{0.4\textwidth}{!}{\includegraphics{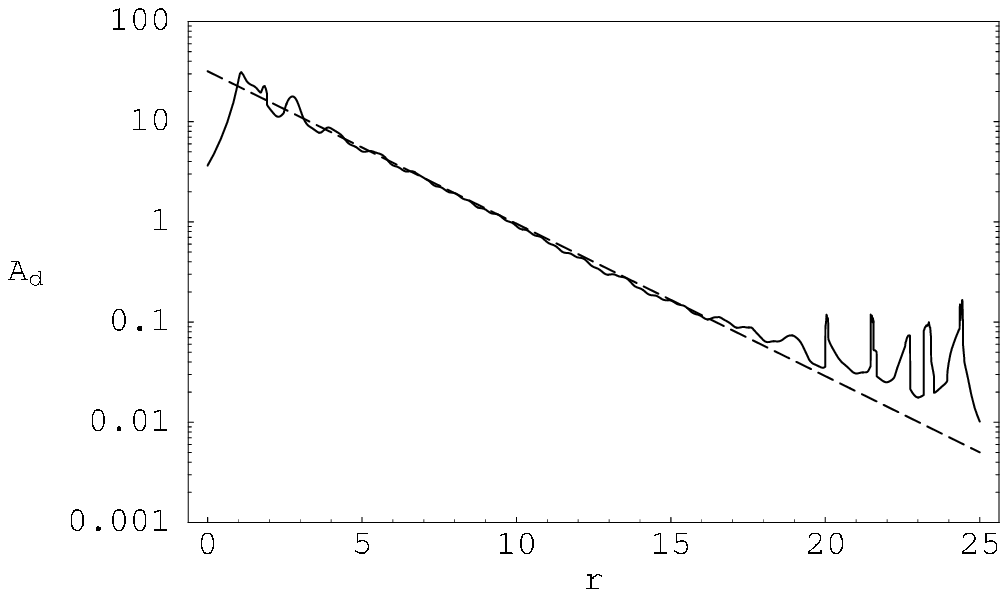}} \par}

\caption{\label{Fig: RRDF WT amplitude}Instantaneous amplitude \protect\( A_{d}(r)\protect \)
of the largest component of the RRDF \protect\( d(r)\protect \) (see
Fig.~\ref{Fig: RRDF}) plotted on a logarithmic scale. Solid line
is \protect\( A_{d}(t)\protect \) extracted using Eq.~(\ref{Eqn: Instantaneous amplitude})
with \protect\( \sigma =2\protect \), dashed line is the best-fit
exponential decay \protect\( A_{d}(r)\simeq \alpha e^{-\beta r}\protect \)
over the range \protect\( 5<r<18\protect \), where \protect\( \alpha =31.8\pm 0.5\protect \)
and \protect\( \beta =0.350\pm 0.001\protect \).}
\end{figure}

The amplitudes of components in an RRDF are expected to tend to zero
in the limit \( r\rightarrow \infty  \) for a disordered material
due to the absence of long-range order. The instantaneous amplitude
\( A_{d}(r) \) of the dominant ridge extracted from the CWT using
the new wavelet, Eq.~(\ref{Eqn: Harrop wavelet}), is shown plotted
on a logarithmic scale in Fig.~\ref{Fig: RRDF WT amplitude}. The
amplitude is clearly seen to decay exponentially in the region \( 2<r<18 \).
The reason for the exponential form of this decay (also observed by
Ding et al.~\cite{Ding1998} for silica glass) is not yet known.

The method used by Ding et al.~\cite{Ding1998} could not reproduce
the frequency modulation of the damped density fluctuations in \( d(r) \)
and their observed amplitude modulation contained only six points
which were noted to decay approximately exponentially. In comparison,
our method reproduces true, instantaneous frequencies (analogous to
frequencies obtained by Fourier analysis), showing the frequency modulation
of individual density fluctuations in \( d(r) \), and the amplitude
modulations of these variable-frequency components, as a continuum
of points. This gives much more compelling evidence for the exponential
decay first observed by Ding et al.~\cite{Ding1998}. In addition,
we can detect a significant deviation from the exponential decay of
\( A_{d}(r) \) at large \( r \) (see Fig.~\ref{Fig: RRDF WT amplitude}).
This may either be due to statistical noise from the finite nature
of the simulation or due to the use of periodic boundary conditions
in the model producing effective long-range order. The precise reason
needs further investigation.

\section{Conclusions\label{Sec: Conclusions}}

We have identified the complex, continuous wavelet transform using
wavelets of constant shape~\cite{Daubechies1992} as a method well
suited to the time-frequency analysis of one-dimensional functions.
For our target application, namely the analysis of functions with
components which have rapidly varying frequency and amplitude modulations,
we have illustrated an important shortcoming of the existing Morlet
wavelet~\cite{Goupillaud1984}, explained the origin of this shortcoming
and proposed a new wavelet which overcomes the problem. In addition,
we have specialized an existing method~\cite{Delprat1991} for extracting
instantaneous frequency and amplitude measurements from signals to
the new wavelet.

Two example functions have been analysed using the new wavelet and
new method of analysis. The first, a simple variable-frequency function,
illustrates the significant improvement of the new wavelet over the
Morlet wavelet and gives numerical evidence that our method of analysis
is accurate. The second is a real-world example of a direct-space
atomic correlation function of a glass which highlights the advantages
of the method over the conventional Fourier transform and greatly
improves upon the single, previous wavelet analysis of such a function
by Ding et al.~\cite{Ding1998}.

We have successfully used the WT to analyze the reduced radial distribution
function (RRDF) of a model glass and can immediately identify previously
undetected features. The dominant component in the RRDF (the damped
extended-range density fluctuations) has a period which rapidly settles
to a constant value. Other components with different frequencies are
present in the RRDF. These oscillations all have approximately exponentially
decaying real-space amplitudes.


\begin{thebibliography}{13}
\expandafter\ifx\csname natexlab\endcsname\relax\def\natexlab#1{#1}\fi
\expandafter\ifx\csname bibnamefont\endcsname\relax
  \def\bibnamefont#1{#1}\fi
\expandafter\ifx\csname bibfnamefont\endcsname\relax
  \def\bibfnamefont#1{#1}\fi
\expandafter\ifx\csname citenamefont\endcsname\relax
  \def\citenamefont#1{#1}\fi
\expandafter\ifx\csname url\endcsname\relax
  \def\url#1{\texttt{#1}}\fi
\expandafter\ifx\csname urlprefix\endcsname\relax\def\urlprefix{URL }\fi
\providecommand{\bibinfo}[2]{#2}
\providecommand{\eprint}[2][]{\url{#2}}

\bibitem[{\citenamefont{Haar}(1910)}]{Haar1910}
\bibinfo{author}{\bibfnamefont{A.}~\bibnamefont{Haar}}, \bibinfo{journal}{Math.
  Ann.} \textbf{\bibinfo{volume}{69}}, \bibinfo{pages}{331}
  (\bibinfo{year}{1910}).

\bibitem[{\citenamefont{Goupillaud et~al.}(1984)\citenamefont{Goupillaud,
  Grossmann, and Morlet}}]{Goupillaud1984}
\bibinfo{author}{\bibfnamefont{P.}~\bibnamefont{Goupillaud}},
  \bibinfo{author}{\bibfnamefont{A.}~\bibnamefont{Grossmann}},
  \bibnamefont{and} \bibinfo{author}{\bibfnamefont{J.}~\bibnamefont{Morlet}},
  \bibinfo{journal}{Geoexploration} \textbf{\bibinfo{volume}{23}},
  \bibinfo{pages}{85} (\bibinfo{year}{1984}).

\bibitem[{\citenamefont{van~den Berg}(1999)}]{vandenBerg1999}
\bibinfo{author}{\bibfnamefont{J.}~\bibnamefont{van~den Berg}},
  \emph{\bibinfo{title}{Wavelets in Physics}} (\bibinfo{publisher}{Cambridge
  University Press}, \bibinfo{address}{Cambridge, England},
  \bibinfo{year}{1999}).

\bibitem[{\citenamefont{Prasad and Iyengar}(1997)}]{Prasad1997}
\bibinfo{author}{\bibfnamefont{L.}~\bibnamefont{Prasad}} \bibnamefont{and}
  \bibinfo{author}{\bibfnamefont{S.~S.} \bibnamefont{Iyengar}},
  \emph{\bibinfo{title}{Wavelet Analysis with applications to Image
  Processing}} (\bibinfo{publisher}{CRC Press}, \bibinfo{address}{Boca Raton,
  FL, USA}, \bibinfo{year}{1997}).

\bibitem[{\citenamefont{Silverman and Vassilicos}(2000)}]{Silverman2000}
\bibinfo{author}{\bibfnamefont{B.~W.} \bibnamefont{Silverman}}
  \bibnamefont{and} \bibinfo{author}{\bibfnamefont{J.~C.}
  \bibnamefont{Vassilicos}}, \emph{\bibinfo{title}{Wavelets: the key to
  intermittent information?}} (\bibinfo{publisher}{Oxford University Press},
  \bibinfo{address}{Oxford, England}, \bibinfo{year}{2000}).

\bibitem[{\citenamefont{Ding et~al.}(1998)\citenamefont{Ding, Nanba, and
  Miura}}]{Ding1998}
\bibinfo{author}{\bibfnamefont{Y.}~\bibnamefont{Ding}},
  \bibinfo{author}{\bibfnamefont{T.}~\bibnamefont{Nanba}}, \bibnamefont{and}
  \bibinfo{author}{\bibfnamefont{Y.}~\bibnamefont{Miura}},
  \bibinfo{journal}{Phys. Rev. B} \textbf{\bibinfo{volume}{58}},
  \bibinfo{pages}{14279} (\bibinfo{year}{1998}).

\bibitem[{\citenamefont{{D}zugutov}(1993)}]{Dzugutov1993}
\bibinfo{author}{\bibfnamefont{M.}~\bibnamefont{{D}zugutov}},
  \bibinfo{journal}{J. Non-Cryst. Sol.} \textbf{\bibinfo{volume}{156}},
  \bibinfo{pages}{173} (\bibinfo{year}{1993}).

\bibitem[{\citenamefont{Daubechies}(1992)}]{Daubechies1992}
\bibinfo{author}{\bibfnamefont{I.}~\bibnamefont{Daubechies}},
  \emph{\bibinfo{title}{Ten Lectures on Wavelets}} (\bibinfo{publisher}{Society
  for Industrial and Applied Mathematics}, \bibinfo{address}{Philadelphia,
  USA}, \bibinfo{year}{1992}).

\bibitem[{\citenamefont{Delprat et~al.}(1991)\citenamefont{Delprat, Escudi\'e,
  Guillemain, Kronland-Martinet, Tchamitchian, and Torr\'esani}}]{Delprat1991}
\bibinfo{author}{\bibfnamefont{N.}~\bibnamefont{Delprat}},
  \bibinfo{author}{\bibfnamefont{B.}~\bibnamefont{Escudi\'e}},
  \bibinfo{author}{\bibfnamefont{P.}~\bibnamefont{Guillemain}},
  \bibinfo{author}{\bibfnamefont{R.}~\bibnamefont{Kronland-Martinet}},
  \bibinfo{author}{\bibfnamefont{P.}~\bibnamefont{Tchamitchian}},
  \bibnamefont{and}
  \bibinfo{author}{\bibfnamefont{B.}~\bibnamefont{Torr\'esani}},
  \bibinfo{journal}{IEEE Trans. Inf. Th.} \textbf{\bibinfo{volume}{38}},
  \bibinfo{pages}{644} (\bibinfo{year}{1991}).

\bibitem[{\citenamefont{Carmona et~al.}(1997)\citenamefont{Carmona, Hwang, and
  Torresani}}]{Carmona1997}
\bibinfo{author}{\bibfnamefont{R.}~\bibnamefont{Carmona}},
  \bibinfo{author}{\bibfnamefont{W.~L.} \bibnamefont{Hwang}}, \bibnamefont{and}
  \bibinfo{author}{\bibfnamefont{B.}~\bibnamefont{Torresani}},
  \bibinfo{journal}{IEEE Trans. Sig. Proc.} \textbf{\bibinfo{volume}{45}},
  \bibinfo{pages}{2586} (\bibinfo{year}{1997}).

\bibitem[{\citenamefont{Carmona et~al.}(1999)\citenamefont{Carmona, Hwang, and
  Torresani}}]{Carmona1999}
\bibinfo{author}{\bibfnamefont{R.}~\bibnamefont{Carmona}},
  \bibinfo{author}{\bibfnamefont{W.~L.} \bibnamefont{Hwang}}, \bibnamefont{and}
  \bibinfo{author}{\bibfnamefont{B.}~\bibnamefont{Torresani}},
  \bibinfo{journal}{IEEE Trans. Sig. Proc.} \textbf{\bibinfo{volume}{47}},
  \bibinfo{pages}{480} (\bibinfo{year}{1999}).

\bibitem[{\citenamefont{Simdyankin et~al.}(2000)\citenamefont{Simdyankin,
  Taraskin, Dzugutov, and Elliott}}]{Simdyankin2000}
\bibinfo{author}{\bibfnamefont{S.~I.} \bibnamefont{Simdyankin}},
  \bibinfo{author}{\bibfnamefont{S.~N.} \bibnamefont{Taraskin}},
  \bibinfo{author}{\bibfnamefont{M.}~\bibnamefont{Dzugutov}}, \bibnamefont{and}
  \bibinfo{author}{\bibfnamefont{S.~R.} \bibnamefont{Elliott}},
  \bibinfo{journal}{Phys. Rev. B} \textbf{\bibinfo{volume}{62}},
  \bibinfo{pages}{3223} (\bibinfo{year}{2000}).

\bibitem[{\citenamefont{Elliott}(1990)}]{Elliott1990}
\bibinfo{author}{\bibfnamefont{S.~R.} \bibnamefont{Elliott}},
  \emph{\bibinfo{title}{Physics of Amorphous Materials 2nd edn.}}
  (\bibinfo{publisher}{Longman}, \bibinfo{address}{London, England},
  \bibinfo{year}{1990}).

\end{thebibliography}
\end{document}